# The Future of Pedestrian–Automated Vehicle Interactions

By Lionel P. Robert Jr.

When crossing the street one of the first things most pedestrians do when they see an oncoming vehicle is make eye contact with the driver. This is one way to ensure that the driver has seen you. Being seen by the driver is important to ensuring that you can cross the street safely. Now imagine doing the same scenario as a pedestrian, only when you attempt to make eye contact with the driver you discover that the vehicle has no driver. Do you cross the street? That situation is likely to become an everyday occurrence with the widespread adaptation of automated vehicles (AVs). The answer to that question cannot be to wait and let the AV pass and then cross the street. To be fully integrated into our society, AVs need to be navigated in much the same way as other vehicles.

As director of the Michigan Automated Vehicle Research Intergroup Collaboration (MAVRIC), I study ways to make pedestrian–AV interactions safer. Pedestrian–AV interaction is a subfield of human–AV interaction, which focuses on the various touch points between AVs and individuals outside the AV. This area of study is particularly important for several reasons. Pedestrians, unlike individuals in the AV, have not made a conscious decision to subject themselves to the AV. Therefore, they are less likely to be familiar with or comfortable with the technology [1]. The interactions between pedestrians and AVs are also quantitatively and qualitatively different. It is also not clear how the results from research on driver/rider interactions with AVs are directly applicable to understanding pedestrian interactions with AVs. Therefore, at MAVRIC we believe that the study of pedestrian–AV interactions is essential to understanding human–AV interaction.

In this article we present and discuss the current research trends in pedestrian–AV interactions and their challenges. Then, we highlight several important areas that are receiving much less attention but are vital to the study of pedestrian–AV interactions.

### Pedestrian–AV Communications

Research on pedestrians' interactions with manually driven vehicles has highlighted the important role of communications between pedestrians and vehicle drivers in ensuring safe interactions [2]. This communication is often done through verbal exchanges, hand gestures or eye contact between pedestrians and vehicle drivers. The removal of the driver presents new challenges to facilitating the communications needed to ensure pedestrian safety [3]. The research on pedestrian–AV communications can be divided into those examining AV-to-pedestrian communications and those examining pedestrian-to-AV communications.

### AV-to-pedestrian Communication

Research on AV communication with pedestrians focuses on leveraging the use of devices on the AV to promote communications with the pedestrians. The most commonly studied devices are light-emitting diode (LED) message boards. These LED message boards are located on various parts of the AV (e.g., side panels, windshields and overhead) [4]. Research is being conducted to determine the best locations for placing the LED boards on AVs. There is also ongoing research on what information these message boards should display. For example, should they display what the AV is currently doing (i.e. stopping) or what the pedestrian should be doing (i.e. cross now) [5]. One of the biggest limitations to the use of LED messages is related to scalability. An LED board might display a message intended for one pedestrian but read by another pedestrian. For example, an AV's LED board might display a message that it is safe for pedestrian "A" to cross but also have the message read by pedestrian "B" whom the AV was unaware of and to whom the AV did not intend to communicate that it was safe to cross. This could result in at least one pedestrian mis-reading the AV's intention. Another example of the scalability problem is the increase in the cognitive load imposed on a pedestrian as the number of AVs with LED boards increases. As the number of AVs that the pedestrian encounters increases, so does the number of potential LED messages to read. A pedestrian reading one message from one AV is certainty manageable but messages from two, three or four become somewhat more difficult. This is especially true when you factor in the habits and behaviors associated with many pedestrians such as text messaging and email reading. In addition, when you couple the first scalability problem with the second scalability problem it becomes easy to see how issues related to scalability can magnify. Scalability problems are not insurmountable, but they do present ongoing challenges with the use of LED boards as a standalone solution.

### Pedestrian-to-AV Communication

The removal of the driver presents another problem — the ability of the pedestrian to communicate with the AV. Common ground or a shared understanding helps to promote communication. One important source of common ground between pedestrians and drivers is based on their shared experiences. In many cases drivers have been pedestrians and pedestrians have at least ridden in a vehicle, if not driven a vehicle. This creates common ground between the driver and the pedestrian, which facilitates communications. However, AVs have not been pedestrians and AVs do not always mimic human drivers in their behavior or decision-making. Both make it difficult for the AV and the pedestrian to establish common ground. Researchers are conducting studies to determine how pedestrians communicate their intention implicitly through their body language and behavior. Models employing machine learning are being developed to teach AVs how to interpret implicit communications from the pedestrians so that they can react to them correctly. However, the dynamic and emergence nature of these interactions makes modeling these interactions particularly challenging. Imagine the AV interpreting the pedestrian's behavior, then reacting to this interpretation, followed by the pedestrian reacting to the AV's reactions, followed by the AV reacting and so forth. Problems associated with scalability are likely to increase the degree of complexity. Modeling these ongoing interactions among multiple AVs and multiple pedestrians can quickly become quite complex. Once again, these challenges are not insurmountable, but they are far from solved.

## Understudied Areas of Pedestrian–AV Interactions

Despite the progress being made in the study of pedestrian–AV interaction, there troublingly remain several areas that are underexplored. Next, we present and discuss each area.

### Becoming more Inclusive

Much if not all the research being conducted in pedestrian–AV interactions assumes that the pedestrian is a fully able-bodied individual. This at best limits the applicability of what we learn and at worst could lead to inaccurate models of human behavior that will be hugely problematic and possibly dangerous going forward. For example, it might be easy to see the potential limitations associated with the use of LED messages for people with visual impairments. Although people with visual impairments might not directly benefit from these LED message boards it not clear whether they would necessarily be hurt by them. However, it might be much more difficult to understand the problems associated with applying models based on fully able humans to interpret and predict the behavior of a pedestrian in a wheelchair crossing the street. The use of such models could lead to potential safety hazards. To be fair, there are those who believe that from an engineering perspective we should address the "general problem" first (i.e. fully able-bodied pedestrians), then move to the "special cases" later. However, those with experience in human-centered design would warn against this approach. Decisions that are made to address the so-called general problem greatly limit our ability to be more inclusive later. Therefore, it is important that researchers on pedestrian–AV interactions begin to be more inclusive with regard to both their target populations and the problems they attempt to address.

### Tapping into the Infrastructure

One area that remains greatly unexplored is the role of the infrastructure. Research on pedestrian–AV interactions typically assumes that the AV is a standalone vehicle. This assumption offloads much if not all the computational requirements to the AV. However, a much more effective approach would involve leveraging the infrastructure to help simplify the interaction challenges and reducing the computational power required by the AV. Currently, when an AV approaches a crosswalk with a traffic light it must rely on its visual sensing to determine the status of the traffic light (red, yellow, green), then determine whether it should stop, and when and how far from the crosswalk it should stop. Weather conditions can reduce the visibility of the traffic light, thereby increasing the AV's decision time, while road conditions (wet or slippery) can increase the distance needed to stop. Both separately and jointly are likely to increase the potential for AV error. Now, imagine if the AV could directly communicate with the traffic light. The traffic light could directly send its status to the AV (red, yellow, green). This would not only reduce potential problems associated with a lack of visibility but because the communication from the traffic light can be sent at distances greater than the visual range the AV would have more time to stop, thereby reducing the potential for AV error.

## Understanding National and Regional Differences

Pedestrian–vehicle interactions are driven by social norms that vary greatly within and across countries. No one disputes that driving norms among New York, Shanghai and New Delhi differ greatly. Neither would anyone dispute the differences, although not always as profound, exist between New York and Boston or among rural, suburban and urban American roadways. However, it is not clear whether such differences are being captured in the study of pedestrian–AV interactions. Yet, the driving social norms are vital to understanding the expectations one has for pedestrians and the AVs that interact with them. These expectations will greatly impact the communication between pedestrians and AVs. Thus, the study of national and regional differences in pedestrian–AV interactions is vital to safe widespread adoption of AVs.

## Conclusion

Despite the recent advances in the study of pedestrian–AV interactions, there are still important challenges. Researchers have been exploring different ways to overcome barriers to effective communications between pedestrians and AVs. Other important areas also remain largely unexplored. Particularly, there is need to become more inclusive with regard to targeted populations, expand the design space to include the infrastructure and begin to understand national and regional differences.